\documentclass[preprint,showpacs,preprintnumbers,amsmath,amssymb,nofootinbib]{revtex4}
\usepackage{amssymb}
\usepackage[mathscr]{eucal}
\usepackage{graphicx}
\def\be{\begin{equation}}
\def\ee{\end{equation}}
\def\ba{\begin{eqnarray}}
\def\ea{\end{eqnarray}}

\def\M{{\cal M}}
\def\Gauss{{\cal C}_{\rm Gauss}}
\def\ham{{\cal C}_{\rm Ham}}
\def\H{{\cal H}}
\def\Ab{{\bar{{\cal A}}_{\rm KS}}}
\def\cyl{{\rm Cyl_{\rm KS}}}
\def\cylstar{{\rm Cyl}^\star_{\rm KS}}
\def\tr{{\rm Tr\,}}

\def\SU{{\rm SU}}

\def\lp{{\ell}_{\rm Pl}}
\def\q{{}^o\!q}
\def\e{{}^o\!e}
\def\w{{}^o\!\omega}

\def\f{\frac}
\def\T{T}
\def\nu{\tau}

\newcommand{\md}{{\mathrm d}}
\newcommand{\vt}{\vartheta}
\newcommand{\vp}{\varphi}
\newcommand{\R}{{\mathbb R}}
\newcommand{\sgn}{{\mathrm{sgn}}}
\def\S{{\mathbb S}}
\def\F{{}^o\!F}

\begin{document}

\preprint{IGPG--05--09/01,AEI--2005--132}
\title{Quantum geometry and the Schwarzschild singularity}
\author{Abhay\ Ashtekar${}^{1,2}$ and Martin Bojowald${}^{2,1}$}
\address{1. Institute for Gravitational Physics and Geometry,\\
Physics Department, Penn State, University Park, PA 16802, USA\\
2. Max-Planck-Institut f\"ur Gravitationsphysik, Albert-Einstein-Institut,
Am M\"uhlenberg 1, D-14476 Potsdam, Germany}

\begin{abstract}

In homogeneous cosmologies, quantum geometry effects lead to a
resolution of the classical singularity without having to invoke
special boundary conditions at the singularity or introduce ad-hoc
elements such as unphysical matter. The same effects are shown to
lead to a resolution of the Schwarzschild singularity. The
resulting quantum extension of space-time is likely to have
significant implications to the black hole evaporation process.
Similarities and differences with the situation in quantum
geometrodynamics are pointed out.

\end{abstract}

\maketitle

\section{Introduction}
\label{s1}

General relativity provides a subtle and powerful interplay
between gravity and geometry, thereby opening numerous
possibilities for novel phenomena. Among the most spectacular of
the resulting conceptual advances are the predictions that the
universe began with a big bang and massive stars can end their
lives as black holes. In both cases, one encounters singularities.
Space-time of general relativity literally ends and classical
physics comes to a halt.

However, general relativity is incomplete because it ignores
quantum effects. It is widely believed that quantum gravity
effects become significant in the high curvature regions that
develop {before} singularities are formed. These are likely to
significantly change the space-time structure, making the
predictions of general relativity unreliable. Hence, real physics
need not stop at the big bang and black hole singularities. While
the classical space-time does end there, quantum space-time may
well extend beyond.

Attempts at extending physics beyond the big-bang singularity date
back at least to the seventies when mini-superspaces were
introduced and quantum cosmology was born (see, e.g., \cite{cm}).
These investigations sparked off new developments in different
directions ranging from foundations of quantum mechanics to the
development of WKB methods to test semi-classicality in quantum
cosmology, to the introduction of novel Euclidean methods to
calculate the appropriate wave function of the universe. This
deepened our understanding of issues related to the quantum
physics of the universe as a whole \cite{jh-lh}. However, in these
approaches the singularity was not generically resolved. A quantum
extension of space-time either required the introduction of new
principles \cite{hh}, or ad-hoc assumptions, such as existence of
matter violating energy conditions already at the classical level,
or of external clocks that remain insensitive to the infinite
curvature encountered at singularities.

Emergence of quantum Riemannian geometry and the associated
mathematical techniques \cite{alrev,crbook,ttrev} have provided a
new approach to revisit the issue of quantum extensions. Within
the mini-superspaces used in quantum cosmology, the big-bang
singularity could be resolved \emph{without} having to introduce
external boundary conditions at the big-bang, or matter/clocks
with unphysical properties \cite{mb3,mb6,abl}. In the homogeneous,
isotropic mini-superspace coupled to a massless scalar field, in
particular, an exhaustive analysis can be carried out \cite{aps}.
At the analytic level, one can introduce an appropriate inner
product on physical states, define Dirac observables and, using
the two, construct semi-classical states. One can then use
numerical methods to examine the nature of the quantum space-time.
If one begins with a semi-classical state at large late times
(`now') and evolves it back in time, it remains semi-classical
till one encounters the deep Planck regime near the classical
singularity. In this regime the quantum geometry effects dominate.
However, the state becomes semi-classical again on the other side;
the deep Planck region serves as a quantum bridge between two
large, classical space-times. The model is too simple to be
applied reliably to the actual universe we live in. However at a
conceptual and mathematical level it demonstrates the new
possibilities, systematically made available by quantum geometry.

It is then natural to ask if a similar resolution of singularities
occurs also in the context of black holes. Now, there is an
extensive literature on the nature of black hole singularities in
the classical theory. In particular, detailed mathematical
analysis of spherical collapse of uncharged matter was carried out
by Christodoulou, Dafermos and others. It has led to the
expectation that, for black holes formed through gravitational
collapse, the singularity would be generically space-like (see,
e.g., \cite{md} and references therein). Therefore, it is natural
to focus on this case first and ask for the nature of quantum
space-time that would result via their resolution.%
\footnote{Indeed, one does not expect quantum gravity effects to
resolve all singularities. If, for example, they led to a
resolution of the time-like singularity of the negative energy
Schwarzschild space-time, energy would be unbounded below in
quantum gravity and the theory would have unphysical features
\cite{hm}. Rather, one expects that there would be \emph{no}
physical states in quantum gravity that would resemble negative
energy Schwarschild geometry at large distances, whence the issue
of `resolution' would simply not arise.}
If the singularity is resolved and {if} the quantum geometry in
the deep Planck regime again serves as a bridge to a large
classical region beyond, there would be no information loss in the
black hole formation and evaporation process \cite{ab2} (see also
\cite{s'thw,sh3}). Pure states in the distant past would evolve to
pure states in the distant future and one would have a space-time
description of the entire process in a quantum mechanical setting.
Thus, a detailed analysis of the fate of black hole singularities
is of considerable importance also for the fundamental issue of
whether the standard unitary evolution of quantum physics has to
be modified in the setting of black holes.

The purpose of this note is to initiate this investigation using
quantum geometry, in the setting of connection dynamics. We will
focus just on the mini-superspace that is appropriate for
describing the geometry interior to the horizon of a Schwarzschild
black hole. Since this mini-superspace is spatially homogeneous
---of Kantowski-Sachs type--- one can take over the techniques
that have been developed in the setting of homogeneous cosmologies
\cite{mb6}. Our main result is that the quantum scalar constraint
is indeed such that the singularity is resolved. The salient
features of our analysis are as follows. First, we use a
\emph{self-adjoint} Hamiltonian constraint. Therefore, our
analysis can serve as the point of departure for the construction
of the physical Hilbert space either via the group averaging
procedure \cite{group} \emph{or} via deparametrization of the
theory as in \cite{aps}. Second, we spell out the symmetry
reduction procedure that leads to the Kantowski-Sachs
mini-superspace \emph{within connection dynamics}. We will see in
sections \ref{s3} and \ref{s4} that the structure of the resulting
phase space is important to the issue of singularity resolution.%
\footnote{A quantization of the Kantowski-Sachs minisuperspace is
available in the geometrodynamical framework, \cite{m}.
Unfortunately, it does not shed light on singularity resolution
because the location of the singularity in the minisuperspace was
misidentified. In the notation used in \cite{m}, the singularity
lies at $(a=\infty, b=0)$ ---not at $(a=0,b=0)$ as assumed there.
Therefore, neither is the inverse volume operator (17) of \cite{m}
bounded or well-defined at the singularity nor does the discrete
equation (23) of \cite{m} enable one to evolve across the
singularity.}
In particular, this space is an extension of the phase space used
in geometrodynamics in that one allows for degenerate triads (and
hence 3-metrics). Thanks to this enlargement, points of the
reduced phase space corresponding to the singularity do not
constitute a boundary, whence the support of the wave function
`beyond singularity' can be interpreted geometrically. However, we
emphasize that the results of this note constitute only initial
steps for a more complete theory, e.g., along the lines of
\cite{aps}.

Our discussion is organized as follows. In section \ref{s2} we
discuss the classical theory of the Kantowski-Sachs
mini-superspace using connection-dynamics. In section \ref{s3} we
analyze the kinematics of its quantum theory and in section
\ref{s4} we present quantum dynamics. Section \ref{s5} provides a
brief summary and directions for future work.

\section{Classical Theory}
\label{s2}

This section is divided into two parts. In the first, we carry out
the Kantowski-Sachs symmetry reduction of vacuum general
relativity and in the second we discuss the structure of the
resulting phase space.

\subsection{Symmetry reduction}
\label{s2.1}

The portion of the space-time interior to the horizon of a
Schwarzschild black hole can be naturally foliated by  3-manifolds
$M$ with topology $\R\times \S^2$ such that the geometry on each
slice is invariant under the Kantowski--Sachs symmetry group
$G=\R\times {\rm SO(3)}$. In the discussion of this model we
closely follow \cite{abl} where the mathematical structure of the
simpler, isotropic model is studied in detail.

As in any homogeneous situation, $M$ acquires an equivalence class
of positive definite metrics related by an overall constant. We
fix one such metric $\q_{ab}$ as well as an orthonormal triad
$\e_i^a$ and co-triad $\w_a^i$, compatible with it. With the given
symmetry, we choose the metric on the orbits of the ${\rm SO(3)}$
action to be the unit 2-sphere metric,
$\md\vt^2+\sin^2\vt\md\vp^2$ in polar coordinates $\vt,\vp$.
Locally, the corresponding co-triad elements $\md\vt$, and
$\sin\vt\md\vp$ can be completed to an orthonormal co-triad by a
third form $\alpha$.  Thanks to the Maurer--Cartan relations for
the symmetry group, $\alpha$ must be closed and therefore exact on
the given topology, whence it defines a third adapted coordinate
$x$ via $\alpha=\md x$. In these coordinates the background metric
has line element
 $$\md s_o^2= \md x^2+\md\vt^2+\sin^2\vt\md\vp^2$$
and determinant $\q=\sin^2\vt$.

In connection dynamics, the canonically conjugate pair consists of
fields $(A_a^i, E^a_i)$ of an ${\rm SU(2)}$ connection 1-form
$A_a^i$ and a (possibly degenerate) triad $E^a_i$ of density
weight 1 on $M$. This pair is invariant under the symmetry group
if it satisfies:
\be \mathcal{L}_\xi\, A = D\Lambda \quad {\rm and} \quad
\mathcal{L}_\xi E = [E, \Lambda] \label{sym} \ee
for any symmetry vector field $\xi^a$ and some generator
$\Lambda^i$ of local, ${\rm SU(2)}$ gauge transformations (which
may depend on $\xi^a$). One can verify that any such symmetric
pair is gauge equivalent to fields of the form:%
 \footnote{More precisely, invariant connections \cite{kn,b}
carry a non-negative, integer-valued topological charge which we
have set equal to one in the above expressions. (In general, it
would multiply the last contribution to (\ref{sym1}); see, e.g.,
\cite{bk}.) Each value of this charge gives rise to an independent
sector of invariant connections. However, only the sector used
here allows non-degenerate triads invariant under (\ref{sym}) with
{\em the same} local gauge transformation in $A$ and $E$. That $A$
in (\ref{sym1}) satisfies (\ref{sym}) can be verified as follows.
If we denote the three generators of the ${\rm SO(3)}$ action on
$M$ by
 $X=\sin\vp\partial_{\vt}+\cot\vt\cos\vp\partial_{\vp}$,
 $Y=-\cos\vp\partial_{\vt}+\cot\vt\sin\vp\partial_{\vp}$, and
 $Z=\partial_{\vp}$, and the generator of the $\R$-action by $t^a$,
then $\mathcal{L}_X A = D(\cos\vp/\sin\vt)\tau_3$,\,
$\mathcal{L}_Y A = D (\sin\vp/\sin\vt)\tau_3$,\, $\mathcal{L}_Z A
= 0$ and $\mathcal{L}_t A = 0$. Invariance of $E$ in (\ref{sym2})
follows analogously, or by noting that $E^a_i\delta A_a^i$ defines
a gauge invariant and (spatially) constant 1-form on phase space.}
\ba A &=& \tilde{c}\tau_3\md
x+(\tilde{a}\tau_1+\tilde{b}\tau_2)\md\vt
+(-\tilde{b}\tau_1+\tilde{a}\tau_2)\sin\vt\md\vp+
\tau_3 \cos\vt \md\vp \label{sym1}\\
E &=& \tilde{p}_c\tau_3\sin\vt\frac{\partial}{\partial
x}+(\tilde{p}_a\tau_1+\tilde{p}_b\tau_2)
\sin\vt\frac{\partial}{\partial\vt}
+(-\tilde{p}_b\tau_1+\tilde{p}_a\tau_2)\frac{\partial}{\partial\vp}\,.
\label{sym2} \ea
where $\tilde{a},\tilde{b},\tilde{c}$ are real constants and
$\tau^j$ is the standard basis in ${\rm su(2)}$, satisfying
$[\tau_i,\, \tau_j] = \epsilon_{ij}{}^k \tau_k$. However, this
does not exhaust the freedom to perform local ${\rm SU(2)}$ gauge
transformations entirely: There is still the freedom to perform a
\emph{global} $\SU(2)$ transformation along $\tau_3$ which rotates
the `vectors' $(\tilde{a}, \tilde{b})$ and $(\tilde{p_a},
\tilde{p_b})$. We will return to this freedom in the next
sub-section.

From the invariant triad density (\ref{sym2}) we can derive the
corresponding co-triad $\omega$:
\be\label{cotriad} \omega = \omega_c \tau_3 \md x +
(\omega_a\tau_1 + \omega_b\tau_2) \md \vt + (-\omega_b \tau_1 +
\omega_a \tau_2) \sin\vt \md\vp\ee
where
\be \omega_c = \frac{{\rm sgn}\tilde{p}_c\,
\sqrt{\tilde{p}^2_a+\tilde{p}^2_b}}{\sqrt{|\tilde{p}_c|}};\quad
 \omega_b = \frac{\sqrt{|\tilde{p}_c|}\, \tilde{p}_b}
{\sqrt{\tilde{p}^2_a+\tilde{p}^2_b}}; \quad {\rm and} \,\,
\omega_a = \frac{\sqrt{|\tilde{p}_c|}\,\tilde{p}_a}
{\sqrt{\tilde{p}^2_a+\tilde{p}^2_b}}\, . \ee
The co-triad in turn determines its spin-connection, the ${\rm
su(2)}$-valued 1-form $\Gamma$. As one might expect from the
structure of the triad, $\Gamma$ turns out to be the standard
`magnetic monopole' connection:
\begin{equation} \label{Gamma}
 \Gamma= \cos\vt\tau_3\, \md\vp\,.
\end{equation}
Consequently, the ${\rm su(2)}$-valued extrinsic curvature 1-form
$K$ is given by:
\begin{equation} \label{K}
 \gamma K : =A - \Gamma= \tilde{c}\tau_3\md x +
(\tilde{a}\tau_1+\tilde{b}\tau_2)\md\vt
 + (-\tilde{b}\tau_1+\tilde{a}\tau_2)\sin\vt\md\vp\,.
\end{equation}
Finally the curvature $F_{ab}$ of $A_a$ can be easily computed.
Its dual, the ${\rm su(2)}$-valued vector density $B^a\, :=\,
\frac{1}{2} \eta^{abc}F_{bc}$ has the same invariant form as $E^a$:
\begin{equation} \label{B}
 B=(\tilde{a}^2+\tilde{b}^2-1)\tau_3\sin\vt\frac{\partial}{\partial
 x}+(\tilde{a}\tau_1+\tilde{b}\tau_2)\tilde{c}
\sin\vt\frac{\partial}{\partial\vt}
 +(-\tilde{b}\tau_1+\tilde{a}\tau_2)\tilde{c}\frac{\partial}{\partial\vp}
\end{equation}
These expressions will be used in sections \ref{s3} and \ref{s4}.

\subsection{The reduced phase space}
\label{s2.2}

The symmetry reduction procedure led us to a 6-dimensional reduced
phase space $\tilde{\bf \Gamma}$ with coordinates $(\tilde{a},
\tilde{b}, \tilde{c};\, \tilde{p}_a, \tilde{p}_b, \tilde{p}_c)$.
Let us begin by computing the symplectic structure $\tilde{\bf
\Omega}$ on it. The basic idea, as in all symmetry reductions, is
to pull-back the symplectic structure of the full theory to the
symmetry reduced phase space. However, since the full symplectic
structure involves an integral over $M$ and since the fields of
interest are homogeneous also in the non-compact
$\mathbb{R}$-direction, as usual, we are led to consider only a
finite interval $\cal{I}$ in the $\mathbb{R}$ direction. Let the
length of this interval (w.r.t. the fiducial metric $\q_{ab}$) be
$L_o$. Then, the symplectic structure on the reduced phase space
is given by:
\begin{equation}
\tilde{\bf \Omega} = \frac{L_o}{2\gamma G}(2\md\tilde{a}\wedge\md
\tilde{p}_a+ 2\md\tilde{b}\wedge\md \tilde{p}_b+ \md
\tilde{c}\wedge\md \tilde{p}_c)
\end{equation}
where $\gamma$ is the Barbero--Immirzi parameter and $G$ the
gravitational constant. (The volume of the $\S^2$ orbits of the
${\rm SO(3)}$ action, defined by the fiducial metric, is the standard
one, $4\pi$.)

Because of the form (\ref{sym1}) and (\ref{sym2}) of invariant
connections and triads, the vector constraint is automatically
satisfied. However, because of the residual, global ${\rm SU(2)}$
gauge freedom mentioned in section \ref{s2.1}, the Gauss
constraint is not. Inserting the invariant connection and triad
into the Gauss constraint we obtain
\begin{equation}
 \Gauss=\tilde{a}\tilde{p}_b-\tilde{b}\tilde{p}_a=0
\end{equation}
which generates simultaneous rotations of the pairs
$(\tilde{a},\tilde{b})$ and $(\tilde{p}_a,\tilde{p}_b)$. Thus,
only the `norms' $\sqrt{\tilde{a}^2+\tilde{b}^2}$,
$\sqrt{\tilde{p}_a^2+\tilde{p}_b^2}$ and the `scalar product'
$(\tilde{a}\tilde{p}_a+\tilde{b}\tilde{p}_b)$ are gauge invariant.
We will fix this gauge freedom as follows. If
$(\tilde{a},\tilde{b})=0$, we rotate the triad components such
that $\tilde{p}_a=0$. Otherwise we rotate $(\tilde{a},\tilde{b})$
such that $\tilde{a}=0$ which implies $\tilde{b}\not=0$. Then,
$\Gauss=0$ implies $\tilde{p}_a=0$. There is still a residual
gauge freedom\,\, $\Pi_b\colon (b, p_b) \rightarrow (-b, -p_b)$\,\,
which changes the signs of $b$ and $p_b$ simultaneously. This is
just the parity transformation in the $p_b$ variable. One can
either retain this freedom and ensure that all the final
constructions are invariant with respect to this `$b$-reflection'
\emph{or} eliminate it by a gauge choice such as $\tilde{p}_b \ge
0$. This gauge choice turns out to be particularly convenient for
classical dynamics. In the quantum theory, on the other hand, it
is more natural to retain the freedom at first and then ask that
physical states be invariant under the parity operator
$\hat{\Pi}_b$ implementing this transformation. Therefore, we will
allow both possibilities. In either case the 4-dimensional phase
space carries a global chart $(\tilde{b},\tilde{c};
\tilde{p}_b,\tilde{p}_c)$ (where $\tilde{p}_b \ge0$ if the
`$b$-parity gauge' is fixed.) These coordinates are subject to the
scalar or the Hamiltonian constraint, discussed below.

We first note that these variables and, because of its explicit
dependence on $L_o$, the symplectic structure $\tilde{\bf \Omega}$
depend on the fiducial metric. It is convenient to remove this
dependence by rescaling the variables in a manner that is
motivated by their scaling properties,
\begin{equation}
 (b,c):=(\tilde{b},L_o\tilde{c})\quad,\quad
 (p_b,p_c):=(L_o\tilde{p}_b,\tilde{p}_c)\,.
\end{equation}
We now have a one-to-one parametrization of the gauge fixed phase
space ${\bf \Gamma} $ by $(b,p_b,c,p_c)$ with symplectic structure
\begin{equation} \label{symp}
{\bf \Omega} = \frac{1}{2\gamma G}(2\md b\wedge\md p_b+\md
c\wedge\md p_c)\,.
\end{equation}
For later purposes let us express the volume of our elementary
cell ${\cal I}\times \S^2$ and areas bounded by our preferred
family of curves as functions on the reduced phase space. The
volume is given by:
\begin{equation} \label{vol}
 V=\int\md^3x\sqrt{\left|\det E\right|}=4\pi L_o
 \sqrt{|\tilde{p}_c|} |\tilde{p}_b| = 4\pi\sqrt{|p_c|} |p_b|\,.
\end{equation}
The three surfaces $S_{x,\vp}$, $S_{x,\vt}$ and $S_{\vt,\vp}$ of
interest are respectively bounded by the interval ${\cal I}$ and
the equator, ${\cal I}$ and a great circle along a longitude, and
the equator and a longitude (so that $S_{\vt,\vp}$ forms a quarter
of the sphere $\S^2$). Their areas are given by:
\be \label {area}
 A_{x, \vt} = 2\pi |p_b|, \quad A_{x, \vp} = 2 \pi |p_b|,
 \quad{\rm and} \quad  A_{\vt,\vp} = \pi |p_c|\,. \ee

\begin{figure}
\begin{center}
\includegraphics[width=3in]{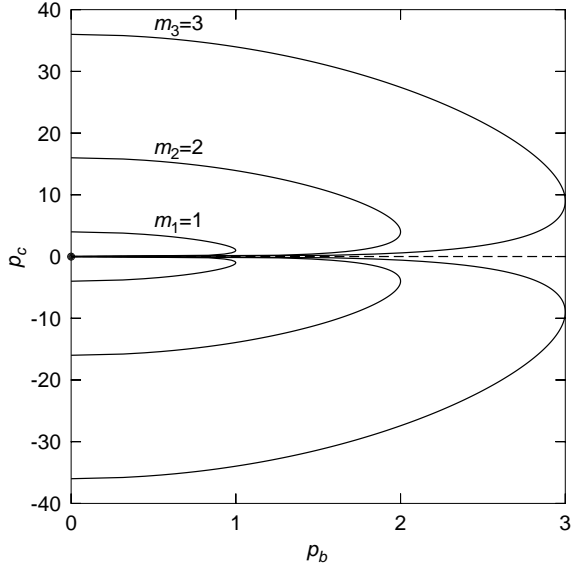}
\caption{\footnotesize Dynamical trajectories in the $p_b-p_c$
plane. Each trajectory reaches the maximum value $m$ of $p_b$ and
meets the $p_c =0$ axis only at the point where $p_b$ also
vanishes (i.e., at the `origin'). Solutions related by the
reflection-symmetry $p_c \rightarrow -p_c$ define the same metric
but carry triads of opposite orientation. For simplicity the
`$b$-parity gauge' is fixed by requiring $p_b \ge 0$.}
\label{traj2}
\end{center}
\end{figure}

Finally, let us consider the Hamiltonian constraint and classical
dynamics. On the reduced phase space ${\bf \Gamma}$, the
constraint functional of the full theory
\begin{equation} \label{fullham}
\ham= \int\md x^3 N e^{-1}\, [\epsilon_{ijk} F_{ab}^i E^a_jE^b_k-
2(1+\gamma^2) K_{[a}^iK_{b]}^j E_i^aE_j^b]\, ,
\end{equation}
where $e:= \sqrt{\left|\det E\right|}\,{\rm sgn}(\det E)$, reduces
to
\begin{eqnarray} \label{ksham}
\ham = -\frac{8\pi N}{\gamma^2}\frac{\sgn p_c}{\sqrt{|p_c|}p_b}
\left[(b^2+\gamma^2)p_b^2+
  2cp_cbp_b\right]\, ,
\end{eqnarray}
where, as is usual in the homogeneous models, we have chosen a
constant lapse function $N$. To simplify the equations of motion,
it is convenient%
\footnote{With this choice, the points $b=0$ or $p_c =0$ have to
be excised in the discussion of dynamics. The explicit form of the
solutions below shows that this is not overly restrictive since on
classical solutions these points correspond to the singularity or
the horizon.}
to choose $N = \frac{\gamma {\rm sgn}p_c\, \sqrt{|p_c|}}{16\pi G
b}$. Then the Hamiltonian constraint becomes $\ham:=-(2\gamma
G)^{-1}[(b^2+\gamma^2)p_b/b+2cp_c]$, yielding the following
equations of motion:
\begin{eqnarray*}
 \dot{b} = -\frac{1}{2}(b+\gamma^2b^{-1}), &\quad\quad&
 \dot{p}_b = \frac{1}{2}(p_b-\gamma^2b^{-2}p_b)\\
 \dot{c} = -2c, &\quad\quad&
 \dot{p}_c = 2p_c\, ,
\end{eqnarray*}
where the `dot' denotes derivative with respect to the affine
parameter of the Hamiltonian vector field. The equations for
$(c,p_c)$ can easily be solved by $c(\T)=c_0e^{-2\T}$ and
$p_c(\T)=p_c^{(0)}e^{2\T}$. For the other components we obtain
$b(\T)=\pm\gamma\sqrt{e^{-(\T-\T_0)}-1}$ for $b$ and, using this
result, $p_b(T)=p_b^{(0)}\sqrt{e^{\T+\T_0} -e^{2\T}}$. If we
introduce the new time parameter $t:=e^{\T}$ and the constant
$m:=\frac{1}{2}e^{\T_0}$, this gives
\begin{eqnarray}
 b(t) = \pm\gamma\sqrt{(2m-t)/t}, &\quad\quad&
 p_b(t) = p_b^{(0)}\sqrt{t(2m-t)}\\
 c(t) = \mp \gamma m p_b^{o}\, t^{-2}\label{3}, &\quad\quad&
 p_c(t) = \pm \,t^2\label{4}
\end{eqnarray}
where we have fixed the multiplicative constant in the expressions of
$p_c(t)$ and $c(t)$, respectively, by requiring that $|p_c|$ be the
geometric radius of the 2-sphere orbits of the ${\rm SO(3)}$ symmetry
and by using the Hamiltonian constraint. Projections to the $p_b-p_c$
plane of typical trajectories are shown in figure \ref{traj2}. It is
clear from the figure that $p_c$ can be taken to be the `internal time
parameter'. In section \ref{s4.1} this interpretation will let us
regard the quantum Hamiltonian constraint as an `evolution equation'
with respect to an internal time parameter defined by the eigenvalues
of the operator $\hat{p}_c$.

Note that we have a 2-parameter family of solutions, labeled by
$m$ and $p^{(0)}_b$, as one would expect from the fact that the
reduced phase space is 2-dimensional. However, in a space-time
description we expect only a 1-parameter family, labeled by $m$.
This discrepancy can be traced back to the standard tension
between the Hamiltonian and space-time notions of gauge (for a
detailed discussion in the context of Bianchi models, see
\cite{as}, and in spherically symmetric models, \cite{kt,k}). In
the Hamiltonian description, rescalings of $p^{(0)}_b$ do not
correspond to gauge because they are not generated by any
constraint. In the space-time description, on the other hand, they
can be absorbed by a rescaling of the coordinate $x$. Therefore,
to make contact with the space-time description, let us fix
$p^{(0)}_b =1$. Then, the above solution to the evolution
equations defines a 1-parameter family of 3-metrics:
\be q_{ab}(t) = \left(\frac{2m}{t} -1\right) \nabla_a x \nabla_b x
+ t^2 (\nabla_a\vt \nabla_b \vt + \sin^2\vt  \nabla_a
\vp\nabla_b\vp ) \ee
defining precisely the Schwarzschild solution of mass $m$,
interior to the horizon. It will be useful to note that, in this
solution $p_c>0$, $p_b=0$ at the horizon ($t=2m$) and $p_c =0$ and
$b,c$ diverge at the singularity.

We will see in the next section that it is convenient to regard
the space $\M$ spanned by $p_b$ and $p_c$ as the mini-superspace
on which quantum wave functions are defined. (In that discussion,
we will not fix the `$b$-parity gauge'; so $p_c$ \emph{and} $p_b$
take values on the entire real line.) Let us therefore interpret
various regions in $\M$. The triad is non-degenerate everywhere
except when $p_b=0$ or $p_c =0$. However, the two degeneracies are
of very different geometric and physical origin. Geometrically
(i.e., independent of classical dynamics) $p_c=0$ separates two
regions with triads of opposite orientations, given by $\sgn\det
E=\sgn(p_c p_b^2)$. On the other hand, as (\ref{cotriad})
(together with our gauge choice which ensures $a=p_a =0$) shows,
the orientation would not have changed even if we had extended
$\M$ across $p_b=0$, allowing for negative values for $p_b$. More
importantly, while the co-triad (\ref{cotriad}) remains smooth and
only becomes degenerate along $p_b=0$, it diverges along $p_c =0$,
signaling the classical singularity. This suggests that the line
$p_b =0$ corresponds to the horizon and $p_c =0$ to the
singularity. This is confirmed from physical considerations
involving classical dynamics. Classical solutions can be smoothly
extended to the line $p_b=0$ which is reached for $t=2m$. The
curvature $B$ remains well-defined there (see (\ref{B})). The
other line, $p_c=0$, is very different. Along classical solutions,
the line $p_c=0$ is approached as $t\to 0$, whence the components
(and invariants) of curvature $B$ also diverge. This line
corresponds to the singularity and cannot be crossed by any
classical dynamical trajectory. Finally, note that since the line
$p_c=0$ separates the two regions of $\M$, where only the triad
orientation is reversed, the two get identified in
geometrodynamics. Therefore the singularity would lie at a
boundary of the mini-superspace in geometrodynamics, making the
geometrical meaning of `evolution across the singularity' obscure.
Furthermore, since the Arnowitt--Deser--Misner (ADM) variables are
based on covariant metrics (rather than contravariant triad
densities), as in other homogeneous models \cite{mb6}, one of the
ADM variables becomes {\em infinite} at the classical singularity.
This further complicates a discussion of the singularity structure
and its resolution.

\section{Quantum Kinematics}
\label{s3}

This section is divided into two parts. In the first we introduce
the basic quantization procedure and in the second we discuss
quantum geometry with emphasis on the classical singularity.

\subsection{Basics}
\label{s3.1}

As in the isotropic model \cite{abl}, we will follow the general
procedure used in the full theory \cite{alrev,crbook,ttrev}. Thus
the elementary configuration variables will be given by holonomies
along curves in $M$ and the momenta by fluxes of triads along
2-surfaces in $M$. However, because of symmetry reduction, one
need not consider all (piecewise analytic) curves or surfaces; a
judiciously chosen, smaller subset suffices to obtain a set of
functions that is sufficiently large to separate points of the
reduced phase space.

Let us begin with the holonomies. We will restrict ourselves to
three sets of curves: those along the $\R$ direction of $M$ with
oriented length $\nu L_o$; those along the equator of $\S^2$ with
oriented length $\mu$; and those along the longitudes of $\S^2$
also with oriented length $\mu$, where all lengths and
orientations are defined using the fiducial triad. (Thus $\nu,
\mu$ are positive if the tangent to the respective curves are
parallel to the triad vectors and negative if they are
anti-parallel.) Holonomies along these curves suffice to
completely determine any invariant connection (\ref{sym1}).%
\footnote{In the fully gauge fixed setting, we could omit
$h_{\vt}$ since it is related to $h_{\vp}$ by conjugation with
$\exp\,\tau_3$. However, we will use a more `democratic' approach
which is applicable also when $a$ is not made to vanish by gauge
fixing.}
\begin{eqnarray}
 h^{(\nu)}_x(A) &=& \exp\int_0^{\nu L_o}\md x \tilde{c}\tau_3=
\cos\frac{\nu c}{2}+ 2\tau_3\sin\frac{\nu c}{2} \label{hol1}\\
h^{(\mu)}_{\vp}(A) &=& \exp-\int_0^{\mu}\md\vp \tilde{b}\tau_1=
\cos\frac{\mu b}{2}- 2\tau_1\sin\frac{\mu b}{2} \label{hol2}\\
h^{(\mu)}_{\vt}(A) &=& \exp\int_0^{\mu}\md\vt \tilde{b}\tau_2=
\cos\frac{\mu b}{2}+ 2\tau_2\sin\frac{\mu b}{2}\, .\label{hol3}
\end{eqnarray}
Matrix elements of these holonomies are functions of the reduced
connection and constitute our configuration variables. Elements of
the algebra they generate are almost periodic functions of $b$ and
$c$ of the form $f(b,c) = \sum_{\mu,\nu}\, f_{\mu\nu} \exp
{\frac{i}{2}\, (\mu b + \nu c)}$, where $f_{\mu,\nu} \in
{\mathbb{C}}$, $\mu,\nu \in \mathbb{R}$ and the sum extends over a
finite set. This algebra is the Kantowski-Sachs analog of the
algebra of cylindrical functions in the full theory
\cite{alrev,crbook,ttrev}. We will therefore denote it by $\cyl$.
Consider the $C^\star$ algebra obtained by completing $\cyl$ using
the sup norm. The quantum configuration space $\Ab$ is the
Gel'fand spectrum of this algebra. From the structure of the
algebra, it follows that the spectrum is naturally isomorphic to
the Bohr compactification $\bar{\R}_{\rm Bohr}^2$ of the Abelian
group $\R^2$ \cite{afw,jw}. (Recall that in the isotropic case
\cite{abl}, the quantum configuration space is isomorphic to
$\bar{\R}_{\rm Bohr}$.) The Hilbert space $\tilde\H$ is obtained
by the Cauchy completion of $\cyl$ with respect to the natural
Haar measure $\mu_o$ on the Abelian group $\bar{\R}^2_{\rm Bohr}$;
$\tilde\H = L^2(\bar{\R}_{\rm Bohr}^2,\md\mu_o)$. Employing the
standard bra-ket notation, we can define a basis $|\mu,\nu\rangle$
in $\tilde\H$ via:
\begin{equation} \label{basis}
 \langle b,c|\mu,\nu\rangle = e^{\frac{i}{2}\,(\mu b+\nu c)} \qquad
 \mu,\nu\in{\R}\, .
\end{equation}
This is an orthonormal basis:
\be \langle \mu^\prime, \nu^\prime|\mu,\nu\rangle =
\delta_{\mu^\prime\, \mu}\, \delta_{\nu^\prime\, \nu} \,  \ee
where, on the right side, we have the Kronecker symbol, rather
than the Dirac delta distribution. Thus, each basis vector is
normalizable and has unit norm.

As one would expect, the configuration ---i.e., holonomy---
operators $\hat{h}_x^{(\nu)},\, \hat{h}_\vp^{(\mu)},\,
\hat{h}_\vt^{(\mu)}$ operate on $\tilde\H$ by multiplication.
However, as in the full theory, these operators are not required
to be weakly continuous in parameters $\mu$ and $\nu$, whence
there are no operators $\hat{b}, \hat{c}$ corresponding to the
connection itself. consequently, although we have only a finite
number of degrees of freedom, the von-Neumann uniqueness theorem
is inapplicable and this quantum theory is inequivalent to a
standard `Schr\"odinger quantization' (for a further discussion,
see \cite{afw,jw}). We will see that, as in the isotropic model
\cite{abl}, this inequivalence has important consequences.

For the momentum operators we consider fluxes of triads along
preferred 2-surfaces. Apart from fixed kinematical factors, they
are given by components $p_b,p_c$ of triads which, in view of the
symplectic structure (\ref{symp}), are represented by operators%
\footnote{In this paper, we use the standard quantum gravity
convention, $\lp^2 = G\hbar$. Unfortunately, this is different
from the convention $\lp^2 = 8\pi G\hbar$ used in most of the loop
quantum cosmology literature. Hence care should be exercised while
comparing detailed numerical factors.}
\be \hat{p}_b = -i{\gamma\lp^2}\,\frac{\partial}{\partial b},
\quad\quad \hat{p}_c = -2i\gamma\lp^2 \frac{\partial}{\partial c}\,.
\ee
%
%
Their eigenstates are the basis states (\ref{basis}),
\be \hat{p}_b|\mu,\nu\rangle =
\textstyle{\frac{1}{2}}\,\gamma\lp^2\, \mu |\mu,\nu\rangle,\qquad
\hat{p}_c|\mu,\nu\rangle = \gamma\lp^2\, \nu
 |\mu,\nu\rangle\, . \ee

However, we still have to incorporate the residual gauge freedom
which corresponds to a parity reflection in the $b$ degree of
freedom. Therefore, only those states in $\tilde\H$ which are
invariant under the parity operator $\hat\Pi_b\colon |\mu,\nu\rangle
\rightarrow |-\mu,\nu\rangle$ can belong to the kinematical
Hilbert space $\H$. A basis in $\H$ is thus given by:
\be \frac{1}{\sqrt{2}}\, [|\mu, \nu\rangle + |-\mu, \nu\rangle
]\ee
Finally, we express the volume operator in terms of triad
operators $\hat{p}_b, \hat{p}_c$. Recall that the region of $M$
under consideration is of the type ${\cal I} \times \S^2$, where
${\cal I}$ has length $L_o$ with respect to the fiducial metric.
From the classical expression (\ref{vol}) of the volume $V$ of
this region, it follows that the operator $\hat{V}$ is given by
$\hat{V}=4\pi |\hat{p}_b|\, \sqrt{|\hat{p}_c|}$. It is diagonal in
our $|\mu, \nu\rangle$ basis and the eigenvalues are:
\begin{equation}
 V_{\mu\nu}=2\pi\, \gamma^{3/2}\, |\mu|\, \sqrt{|\nu|}\,\, \lp^3\, .
\end{equation}
As in the isotropic case, the volume of our cell ${\cal I} \times
\S^2$ with respect to the fiducial metric can be large, but its
volume in the `elementary' state $\exp \frac{i}{2}\,(b+c)$, is of
Planck size.

\subsection{Quantum geometry}
\label{s3.2}

As in the isotropic case the triad operators $\hat{p}_b$,
$\hat{p}_c$ commute whence, in contrast to the full theory, one can
construct the triad representation. In this sub-section we will
study the quantum Riemannian geometry using this representation.
In the next sub-section we will see that the representation is
also convenient for dynamics, for it suggests an intuitive
interpretation for the action of the Hamiltonian constraint.

Let us then expand a general state in terms of eigenstates of the
triad operators, $|\psi\rangle= \sum_{\mu\nu}\psi_{\mu\nu}
|\mu,\nu\rangle$, and use the coefficients $\psi_{\mu\nu}$ to
represent the state. This wave function $\psi_{\mu\nu}$ is
supported on the mini-superspace $\mathcal{M}$ coordinatized by
$p_b$ and $p_c$. In the classical theory, already at the kinematical
level, the lines $p_b =0$ and $p_c =0$ are special: the co-triad
(\ref{cotriad}) becomes degenerate along $p_b=0$ (and the line
represents the horizon) while it diverges on the line $p_c=0$ (and
this line represents the singularity). It is then natural to
examine wave functions $\psi_{\mu\nu}$ which are supported on
lines $\mu=0$ and $\nu=0$ and ask, already at the kinematical
level, for the nature of quantum geometry they represent.

Let us consider the co-triad  component $\omega_c \equiv
\sgn(p_c)\, |p_b|/\sqrt{|p_c|}$ along $\md x$ (see
(\ref{cotriad})). Since it involves an inverse power of $p_c$ and
since there exist normalizable kets $|\mu, \nu=0\rangle$ of the
operator $\hat{p}_c$ with zero eigenvalue, the naive operator
obtained by replacing $p_b$ and $p_c$ by corresponding operators
is not even densely defined, let alone self-adjoint. To define
$\hat\omega_c$, therefore, a new strategy is needed. We will
follow what is by now a standard procedure, adapted from
Thiemann's analysis in the full theory \cite{tt}. The first step
is to express $\omega_c$ in terms of the elementary variables
which do have unambiguous quantum analogs ---holonomies and
positive powers of $p_b, p_c$--- and Poisson brackets between
them. On the classical phase space, we have the exact equality:
\begin{equation} \label{omegaccomm}
 \omega_c=\frac{1}{2\pi\gamma
G}\tr\left(\tau_3 h_x \{h_x^{-1},\,V\}\right)
\end{equation}
%
%
%
where $h_x$ is the holonomy along the interval ${\cal I}$, i.e.,
along the edge of length $L_o$ (with respect to the fiducial
metric) we fixed in our construction of the phase space. Then,
replacing $h_x$ and $V$ by their unambiguous quantum analogs and
the Poisson bracket by $1/i\hbar$ times the commutator, we obtain
\begin{eqnarray}
 \hat{\omega}_c &=& -\frac{i}{2\pi \gamma\lp^2}\, \tr\left(\tau_3
\hat{h}_x[\hat{h}_x^{-1},\, \hat{V}]\right)\nonumber\\
 &=& -\frac{i}{2\pi \gamma\lp^2}\left(\sin\frac{
c}{2}\hat{V}\cos\frac{c}{2}- \cos\frac{c}{2}\hat{V}\sin\frac{
c}{2}\right)\,.
\end{eqnarray}
%
%
It turns out that this operator is diagonalized by our basis
$|\mu, \nu\rangle$ of (\ref{basis}),
\begin{equation}
\hat{\omega}_c |\mu,\nu\rangle = \frac{1}{4\pi \gamma\lp^2}
(V_{\mu,\nu+ 1}-V_{\mu,\nu- 1})\, |\mu,\nu\rangle =
\frac{\sqrt{\gamma}}{2}\lp\, |\mu|\, (\sqrt{|\nu+ 1|}-\sqrt{|\nu-
1|})\,\, |\mu,\nu\rangle\,.
\end{equation}

For large $\nu$, far away from the classical singularity, these
eigenvalues are very close to the classical expectation
$|\omega_c|= |p_b|/\sqrt{|p_c|}$. This is even true for small
$\mu$, i.e.\ we do not need to be far from the horizon. Closer to
the classical singularity $p_c=0$, on the other hand, the behavior
becomes significantly different from the classical one. At the
classical singularity itself, where $|p_b|/\sqrt{|p_c|}$ diverges,
the eigenvalue of $\hat{\omega}_c$ is zero.%
\footnote{Exact classical identities such as (\ref{omegaccomm})
for objects containing inverse powers are available only in
homogeneous models. More generally, the bracket between holonomies
and volume is more complicated \cite{mb10}. These complications
lead to additional correction terms and the resulting operator is
no longer bounded on eigenspaces of $\hat{p}_b$.}

\emph{Remark:} In the above definition of $\hat{\omega}_c$ we
needed a holonomy in the $x$-direction. We chose to evaluate it
using the interval ${\cal I}$ of length $L_o$ with respect to the
fiducial metric. While this choice is natural because this
interval appears already in the construction of the phase space,
we could have used an interval of length $L_o\delta$ for some
$\delta$. Then the resulting operator would also have been bounded
on the entire Hilbert space $\tilde\H$,\, and $|\mu, \nu
=0\rangle$ would again have been eigenvectors of the operator with
zero eigenvalue. Thus, the qualitative properties of the operator
are insensitive to $\delta$. Furthermore, from general
considerations one can argue \cite{abl} that $\delta$ should be of
the order of $1$. However, the value of the upper bound of the
spectrum does depend on $\delta$. Therefore, as in the isotropic
model, the precise numerical coefficient in this bound should not
be attributed physical significance.

Let us now turn to the second triad component $\omega_b = {\rm
sgn} p_b\,\, \sqrt{|p_c|}$. Since $\sqrt{|\nu|}$ is a well-defined
function of $\nu$ on the entire spectrum of $\hat{p}_c$, we can
quantize $\omega_b$ directly. This operator is again diagonal in
our basis $|\mu, \nu\rangle$ and its eigenvalues are given by
\begin{equation}
 \hat{\omega}_b|\mu,\nu\rangle=
\sqrt{\gamma}\lp\sgn(\mu)\sqrt{|\nu|}\,|\mu,\nu\rangle\,.
\end{equation}
Since the operator is quantized directly, the eigenvalues are just
the ones one would expect from the classical expression
(\ref{cotriad}) of $\omega_b$.

The properties of co-triad operators suggest that the lines
$\mu=0$ and $\nu=0$ have very different features also in quantum
geometry. No quantum effects are manifest at $\mu=0$ which
corresponds to the classical horizon. Near and on the line $\nu=0$
which represents the singularity, on the other hand, quantum
effects are large. In particular, they remove the classical
singular behavior of the co-triad. We emphasize, however, that
even in homogeneous models, the ultimate test as to whether or not
a singularity persists upon quantization can only come from
studying the dynamics. The key questions are: Is quantum dynamics
well-defined and deterministic across the classical singularity?
And, does this come about without significant quantum corrections
to geometry near the horizon? Only quantum dynamics will tell us
if the indication provided by the properties of the co-triad
operators are borne out.

\section{Quantum Dynamics}
\label{s4}

This section is divided into three parts. In the first we present
the strategy, in the second we obtain the quantum Hamiltonian
constraint, and in the third we use it to discuss the consequent
resolution of the classical singularity.

\subsection{Strategy} \label{s4.1}

To bring out the similarities and differences between the reduced
model and the full theory, it is instructive to begin with the
Hamiltonian constraint (\ref{fullham}) of the full theory:
\begin{equation}
\ham= \int\md^3 x N e^{-1}\, [\epsilon_{ijk} E^{ai} E^{bj}\,
F_{ab}^k- 2(1+\gamma^2) E_i^aE_j^b\, K_{[a}^iK_{b]}^j ] \, .
\end{equation}
We begin by noting some simplifications that arise because of
spatial homogeneity. Recall that the connection $A$ is related to
the spin-connection $\Gamma$ defined by the triad and the
extrinsic curvature $K$ via $A= \Gamma + \gamma K$. As remarked in
section \ref{s2.1}, symmetries of the model imply that
 $\Gamma$ is the standard magnetic monopole
connection; it does not depend on the phase space point under
consideration. From its expression (\ref{Gamma}), it follows that
its curvature $\Omega$ is given by:
\begin{equation}
\Omega = -\sin\vt\,\tau_3\, \md\vt \wedge \md\vp
\end{equation}
Since it is a `c-number' it can be trivially taken over to quantum
theory. Now, the curvature $F$ of the full connection $A$ can be
expanded out as:
\be F_{ab} = 2\partial_{[a}A_{b]} + [A_a,\, A_b] = \Omega_{ab} +
2\gamma\partial_{[a}K_{b]} + \gamma^2 [K_a,\, K_b] + \gamma [\Gamma_a,\,
K_b]- \gamma [\Gamma_b,\, K_a]\ee
Using the expressions (\ref{Gamma}), (\ref{K}) and (\ref{sym2}) of
$\Gamma$, $K$ and $E$, it is straightforward to verify that:
\ba
 [\Gamma, \, \gamma K ] = b \tau_2 \cos \vt \md\vt\wedge \md\vp
&=& -\gamma \md K\nonumber\\
\epsilon_{ijk} (\partial_{[a} K_{b]}^i) \,E^a_j E^b_k &=& 0
\label{simplify} \ea
These equations now imply that the integrand of the Hamiltonian
constraint can be written simply as%
\footnote{This simplification has a natural origin. Because of
(\ref{simplify}), the right hand side is just
$\epsilon_{ijk}E^{ai} E^{bj}\, {}^+\!F_{ab}^k$, where ${}^+\!F$ is
the curvature of the self-dual connection ${}^+\!A = \Gamma + i
K$.}
\be \ham(x) = e^{-1}\,E^{ai} E^{bj}\, (\epsilon_{ijk} F_{ab}^k-
2(1+\gamma^2) K_{[a}^iK_{b]}^j)  = e^{-1}\,E^{ai} E^{bj}\,
(\epsilon_{ijk}\Omega_{ab}^k - 2 K_{[ai} K_{b]j})\, . \ee
Since $\Omega$ is constant on the entire phase space, the
non-trivial part of the Hamiltonian constraint is thus contained
just in a term quadratic in extrinsic curvature.

To pass to the quantum theory, we have to express the right hand
side in terms of holonomies and triads. There are two possible
avenues.\\
i) Using (\ref{simplify}), the extrinsic curvature terms in the
constraint function $\ham(x)$ can be written in terms of
curvatures $F$ and $\Omega$ so that we have:
\be \label{hami}
 \ham(x) = \frac{1}{\gamma^2}\, \epsilon_{ijk}\, e^{-1}\,
E^{ai}E^{bj}\, [(1 +\gamma^2)\, \Omega_{ab}^k - F_{ab}^k]\, .
\ee
The idea now is to use the discussion of co-triads in section
\ref{s3.2} to obtain the operator analog of $\epsilon_{ijk} \,
e^{-1}\,E^{ai}E^{bj}$; carry $\Omega$ to quantum theory trivially
and use holonomies to express the field strength $F$ as in the
full theory.\\
ii) Alternatively, since we have completely fixed the gauge
freedom to perform internal $\SU(2)$ rotations, we can regard $K$
itself as a connection. It is a direct analog of the connection
$A$ in the spatially flat cosmologies where $\Gamma$ vanishes.
Therefore, we will denote its curvature by $\F$; $\F = \md K + [K,\,
K]$. Then, the constraint functional can also be expressed as:
\be \label{hamii}
 \ham(x) = \frac{1}{\gamma^2}\, \epsilon_{ijk}\,e^{-1}\,
E^{ai}E^{bj}\, [\gamma^2\, \Omega_{ab}^k - \F_{ab}^k] \ee
Now, one can again use the discussion of co-triads in section
\ref{s3.2} to obtain the operator analog of $\epsilon_{ijk}\,
e^{-1}\,E^{ai}E^{bj}$; carry $\Omega$ to quantum theory trivially
and use holonomies to express the field strength $\F$. However,
now holonomies have to be constructed using the connection $K$
rather than $A$. Since our Hilbert space $\H$ is built from
holonomies of $A$, at first it seems difficult to express
holonomies of $K$ as operators on $\H$. However, because of
spatial homogeneity, we can calculate $\ham(x)$ at any point. Let
us choose a point on the equator and use holonomies along the
three sets of curves introduced in (\ref{hol1})--(\ref{hol3}).
Using the expressions (\ref{sym1}) and $(\ref{K})$ of $A$ and
$K$, it follows that, along these curves the holonomies of $A$ and
$K$ are equal! Therefore, the required holonomies of $K$ can
indeed be expressed as operators on $\H$.

Both these strategies are viable and, since they begin with
\emph{exact} expressions (\ref{hami}) and (\ref{hamii}) of
classical constraints, the difference in the resulting operators
is just a quantization ambiguity. The first strategy is more
natural in the sense that it does not require the introduction of
a new connection. On the other hand, since the new connection $K$
is the direct analog of the connection used in spatially flat
cosmologies (in particular in \cite{abl}), to facilitate
comparisons, the second strategy has been used in general,
homogeneous models \cite{mb6,bdv}. For definiteness, we will use
the second strategy in the main discussion and comment at the end
on how the first strategy modifies the final constraint
operator. The analysis of the resolution of the singularity can be
carried out with either methods and the conclusion is the same.

\subsection{Quantum Hamiltonian Constraint} \label{s4.2}

Starting from the classical Hamiltonian constraint
\be \label{ham} \ham = \frac{1}{\gamma^2}\, \int \md^3 x\,\,
\epsilon_{ijk}\, e^{-1}\, E^{ai}E^{bj}\, [\gamma^2\, \Omega_{ab}^k
- \F_{ab}^k]\, ,\ee
where the integral is taken over the elementary cell, we wish to
pass to the quantum operator. The overall strategy is the same as
in \cite{abl,mb6,bdv} and the subtleties are discussed in detail
in \cite{abl}. Therefore, here we will present only the main steps
and comment on differences from the isotropic case treated in
\cite{abl}.

In the isotropic case, the elementary cell is a cube of length
$L_o$ with respect to the fiducial metric. To express the
curvature of the connection and the co-triad operator in the
Hamiltonian constraint, one uses holonomies along the curves of
variable length $\mu_oL_o$ along the edges of this cell. For each
value of $\mu_o$ one obtains a quantum constraint operator
$\hat{{\cal C}}_{\rm Ham}^{(\mu_o)}$. As discussed in \cite{abl},
considerations from the full theory imply that $\mu_o$ should not
be regarded as a regulator; rather, the $\mu_o$-dependence of this
operator should be regarded as a quantization ambiguity. This
ambiguity could not be fixed within the reduced model itself. But
by making an appeal to results from the full theory, one can fix
the value of $\mu_o$ using the minimum eigenvalue of the area
operator in the full theory. (Recall that area enclosed by a loop
enters while expressing curvature in terms of holonomies.)

In the present, Kantowski-Sachs model, the overall procedure is
the same but minor adjustments are necessary because of lack of
isotropy. Since $\S^2$ is compact, our elementary cell $\S^2\times
{\cal I}$ has a geometric edge only along the ${\cal I}$
direction. As in section \ref{s3.1}, we will supplement it with
edges along a longitude and the equator of $\S^2$. Holonomies
along these three edges are then given by
(\ref{hol1})--(\ref{hol3}). Now consider curves of length $L_o\delta$
along ${\cal I}$ and of length $\delta$ each along the equator and the
longitude of $\S^2$. Then, the co-triad function in the
Hamiltonian (\ref{ham}) can be expressed as:
\begin{equation}\label{cotraid}
\epsilon_{ijk}\tau^i\, e^{-1}E^{aj}E^{bk}= -2(8\pi\gamma G{\cal
L}_{(k)})^{-1}\, \epsilon^{abc}\;\,\w_c^k\,
h_k^{(\delta)}\{h_k^{(\delta)-1},V\}
\end{equation}
where on the right hand side the index $k$ runs over $x,\vt,\vp$
and is summed over; $h_k$ is the holonomy along the edge $k$; and
${\cal L}_x = L_o\delta$ and ${\cal L}_\vt = {\cal L}_\vp = \delta$. Note
that this is an exact equality for any choice of $\delta$. Components
of the curvature $\F$ can also be expressed in terms of these
holonomies. Set
\begin{equation}
 h^{(\delta)}_{ij} = h_i^{(\delta)}h_j^{(\delta)}(h_i^{(\delta)})^{-1}
(h_j^{(\delta)})^{-1}
\end{equation}
where $i,j,k$ run over $x,\vt,\vp$. Then, using
(\ref{hol1})--(\ref{hol3}) it is straightforward to verify that
\begin{equation} \label{F}
\F_{ab}^i(x)\tau_i=\frac{\w^i_a\w^j_b}{{\cal A}_{(ij)}}\,\,
(h^{{\cal A}_{(ij)}}_{ij}-1)\, +\,\,O((b^2+c^2)^{3/2}\sqrt{\cal
A})
\end{equation}
where ${\cal A}_{x\vt}=\delta^2L_o={\cal A}_{x\vp}$ and ${\cal
A}_{\vt\vp}=\delta^2$. For $\F_{x,\vt}$ and $\F_{x,\vp}$, (\ref{F}) is
the standard geometric relation between holonomies around closed
loops, the area they enclose and curvature. For $\F_{\vt,\vp}$, on
the other hand, while (\ref{F}) continues to hold, the standard
geometrical interpretation is no longer available because the
edges of length $\delta$ along the equator and a longitude fail to
form a closed loop. Nonetheless, the standard relation is
meaningfully extended because of spatial homogeneity, $\delta^2$
playing the role of the `effective area' ${\cal A}_{\vt,\vp}$.

We now have all the ingredients, $\F$, $\Omega$ and the triad
components, to rewrite the classical constraint (\ref{ham}) in a
way which is suitable for quantization. For the $\F$-term in
(\ref{ham}) we obtain
\begin{eqnarray}\label{Fterm}
 &-&\gamma^{-2}\int\md^3x\, \epsilon_{ijk}\, \F_{ab}^i\,\,
 e^{-1}E^{aj}E^{bk}
 = 2\gamma^{-2}\int\md^3x\,\, \tr\left( \F_{ab}^i\tau_i\;
\epsilon_{jkl}\, e^{-1} E^{aj}E^{bk}\tau^l\right)\nonumber\\
 &=& -4(8\pi\gamma^3G)^{-1}\int\md^3x
\det\w\;\tr\left(\epsilon^{ijk}{\cal A}_{(ij)}^{-1}
(h_{ij}^{(\delta)}-1){\cal L}_{(k)}^{-1}h_k^{(\delta)}
\{h_k^{(\delta)-1},V\}\right) \,+O((b^2+c^2)^{3/2}\delta) \nonumber\\
&=& \left[-16\pi(8\pi\gamma^3G\delta^3)^{-1}\sum_{ijk}\epsilon^{ijk}
\tr\left(h_{ij}^{(\delta)}h_k^{(\delta)} \{h_k^{(\delta)},V\}\right)\right]\,
+O\left((b^2+c^2)^{3/2}\delta\right)\, .
\end{eqnarray}
Next, using $\Omega = -\sin\vt\md\vt\wedge\md\vp \tau_3$, the
$\Omega$ term in (\ref{ham}) becomes:
\begin{eqnarray}\label{Omegaterm}
\int\md^3x\,\epsilon_{ijk}\, \Omega_{ab}^i\,\,
e^{-1}E^{aj}E^{bk}\,\, \; &=& -2\int\md^3x\,
\tr\left(\Omega_{ab}^i\tau_i\;
\epsilon_{jkl}\, e^{-1}E^{al}E^{bk}\tau^l\right)\\
&=& -8(8\pi\gamma G\delta)^{-1}\int\md^3x\sin\vt\;
\tr\left(\tau_3\;{\cal L}_x^{-1}h_x^{(\delta)}\{h_x^{(\delta)-1},V\}\right)\\
&=& -32\pi(8\pi\gamma G\delta)^{-1}\;
\tr\left(\tau_3\;h_x^{(\delta)}\{h_x^{(\delta)-1},V\}\right)\,.
\end{eqnarray}
Set
\be C^{(\delta)} = \big[-2 (\gamma^3
G\delta^3)^{-1}\,\sum_{ijk}\epsilon^{ijk}
\tr\left(h_{ij}^{(\delta)}h_k^{(\delta)} \{h_k^{(\delta)},V\}\right)\big] -
\big[-4(\gamma G \delta)^{-1}
\tr\left(\tau_3\;h_x^{(\delta)}\{h_x^{(\delta)-1},V\}\right)\, \big]\ee
Then the classical Hamiltonian constraint is given by
\be \ham = \lim_{\delta \to 0}\, C^{(\delta)}  \ee
Since all terms in $C^{\delta}$ are expressed purely in terms of our
elementary variables ---holonomies and triads--- which have direct
operator analogs, passage to quantum theory is now
straightforward. We obtain:
\begin{eqnarray}
 \hat{C}^{(\delta)} &=& 2i
(\gamma^3\delta^3\lp^2)^{-1} \tr\left(\sum_{ijk}
\epsilon^{ijk}\hat{h}_i^{(\delta)} \hat{h}_j^{(\delta)} \hat{h}_i^{(\delta)-1}
\hat{h}_j^{(\delta)-1}\hat{h}_k^{(\delta)}
[\hat{h}_k^{(\delta)-1},\hat{V}]+2\gamma^2\delta^2\tau_3 \hat{h}_x^{(\delta)}
[\hat{h}_x^{(\delta)-1},\hat{V}]\right)\nonumber\\\nonumber &=&
4i(\gamma^3\delta^3\lp^2)^{-1}\left(
 8\sin\frac{\delta b}{2}\cos\frac{\delta b}{2} \sin\frac{\delta
c}{2}\cos\frac{\delta c}{2}
 \left(\sin\frac{\delta b}{2}\hat{V}\cos\frac{\delta b}{2}-
 \cos\frac{\delta b}{2}\hat{V}\sin\frac{\delta b}{2}\right)\right.\\
 && + \left.\left(4\sin^2\frac{\delta b}{2}\cos^2\frac{\delta
b}{2}+\gamma^2\delta^2\right)
 \left(\sin\frac{\delta c}{2}\hat{V}\cos\frac{\delta c}{2}-
 \cos\frac{\delta c}{2}\hat{V}\sin\frac{\delta c}{2}\right)\right)\label{C}
\end{eqnarray}
(This is a special case of Eq. (26) in \cite{bdv}.)

 The action of this operator on the eigenstates
$|\mu,\nu\rangle$ of $\hat{p}_b$ and $\hat{p}_c$ is given by
\begin{eqnarray}
 \hat{C}^{(\delta)} |\mu,\nu\rangle &=&
(2\gamma^3\delta^3\lp^2)^{-1}
\left[ 2(V_{\mu+\delta,\nu}-V_{\mu-\delta,\nu})\right.\\
&&\times(|\mu+2\delta,\nu+2\delta\rangle- |\mu+2\delta,\nu-2\delta\rangle-
|\mu-2\delta,\nu+2\delta\rangle+ |\mu-2\delta,\nu-2\delta\rangle)\nonumber\\
&&+\left.(V_{\mu,\nu+\delta}-V_{\mu,\nu-\delta}) (|\mu+4\delta,\nu\rangle-
2(1+2\gamma^2\delta^2)|\mu,\nu\rangle+ |\mu-4\delta,\nu\rangle)\right]\,.
\nonumber
\end{eqnarray}

However, while the classical constraint is a real function on the
kinematical phase space ${\bf \Gamma}$, $\hat{C}^{\delta}$ fails to be
self-adjoint on the kinematical Hilbert space $\H$. We therefore
need to add to it its Hermitian adjoint.%
\footnote{Although the self-adjoint form of the Hamiltonian
constraint was discussed briefly in \cite{mb7}, this point was
ignored in the detailed treatment of \cite{abl}. A detailed
treatment of the self-adjoint constraint and its semi-classical
implications can be found in \cite{jw,abw}. It is used in a
crucial way to obtain the physical Hilbert space in \cite{aps}.}
The result is an operator $\hat{C}_{\rm grav}^{(\delta)}:=\frac{1}{2}
(\hat{C}^{(\delta)}+\hat{C}^{(\delta)\dagger})$ given by:
\begin{eqnarray}
 \hat{C}_{\rm grav}^{(\delta)} |\mu,\nu\rangle &=&
(2 \gamma^3\delta^3\lp^2)^{-1}\,
\left[(V_{\mu+\delta,\nu}-V_{\mu-\delta,\nu}+V_{\mu+3\delta,\nu+2\delta}-
V_{\mu+\delta,\nu+2\delta}) |\mu+2\delta,\nu+2\delta\rangle \right.\nonumber\\
&&- (V_{\mu+\delta,\nu}-V_{\mu-\delta,\nu}+V_{\mu+3\delta,\nu-2\delta}-
V_{\mu+\delta,\nu-2\delta}) |\mu+2\delta,\nu-2\delta\rangle \nonumber\\
&&-(V_{\mu+\delta,\nu}-V_{\mu-\delta,\nu}+V_{\mu-\delta,\nu+2\delta}-
V_{\mu-3\delta,\nu+2\delta}) |\mu-2\delta,\nu+2\delta\rangle \nonumber\\
&&+ (V_{\mu+\delta,\nu}-V_{\mu-\delta,\nu}+V_{\mu-\delta,\nu-2\delta}-
V_{\mu-3\delta,\nu-2\delta}) |\mu-2\delta,\nu-2\delta\rangle \nonumber\\
&&+{\textstyle\frac{1}{2}}(V_{\mu,\nu+\delta}-V_{\mu,\nu-\delta}+V_{\mu+4\delta,\nu+\delta}-
V_{\mu+4\delta,\nu-\delta}) |\mu+4\delta,\nu\rangle \nonumber\\
&&-(1+2\gamma^2\delta^2)(V_{\mu,\nu+\delta}-V_{\mu,\nu-\delta}) |\mu,\nu\rangle
\nonumber\\
&&+\left.{\textstyle\frac{1}{2}}
(V_{\mu,\nu+\delta}-V_{\mu,\nu-\delta}+V_{\mu-4\delta,\nu+\delta}-
V_{\mu-4\delta,\nu-\delta}) |\mu-4\delta,\nu\rangle\right]\,.
\end{eqnarray}

Physical states in quantum theory are those which are symmetric
under the `parity operator' $\hat{\Pi}_b$ and lie in the kernel of
the operator $\hat{C}_{\rm grav}^{(\delta)}$. That is, the only
non-trivial quantum Einstein's equations are:
\be \label{qee1}  (\Psi| \hat{\Pi}_b = 0 \quad,\quad (\Psi|
\hat{C}_{\rm grav}^{(\delta)} = 0 \ee
where $(\Psi|$ is an element of the dual $\cylstar$ of the space
$\cyl$ of finite linear combinations of almost periodic functions
of $b,c$. Let us expand $(\Psi|$ using eigenbras $(\mu,\nu|$ (in
$\cylstar$) of the triad operators:
\be (\Psi| = \sum_{\mu,\nu} \, \psi_{\mu,\nu}\, (\mu,\nu| \ee
Then, $\psi_{\mu,\nu}$ can be regarded as wave functions in the
triad (or, Riemannian geometry) representation. To exhibit the
action of $\hat{C}_{\rm grav}^{(\delta)}$ on $\psi_{\mu,\nu}$ we are
led to separate the cases $\mu \ge 4\delta$ from $\mu < 4\delta$
because of the absolute-value $|\mu|$ in the volume eigenvalues.
For $\mu \ge 4\delta$, quantum Einstein's equation (\ref{qee1})
becomes
\begin{eqnarray} \label{qee2}
\hat{C}_{\rm grav}^{(\delta)}\, \psi_{\mu,\nu}
&=&2\delta(\sqrt{|\nu+2\delta|}+\sqrt{|\nu|})
\left(\psi_{\mu+2\delta,\nu+2\delta}- \psi_{\mu-2\delta,\nu+2\delta}\right)
\nonumber\\
&& +(\sqrt{|\nu+\delta|}-\sqrt{|\nu-\delta|})
\left((\mu+2\delta)\psi_{\mu+4\delta,\nu}-
(1+2\gamma^2\delta^2)\mu\psi_{\mu,\nu}+
(\mu-2\delta)\psi_{\mu-4\delta,\nu}\right)\nonumber\\
&&+2\delta(\sqrt{|\nu-2\delta|}+\sqrt{|\nu|})
\left(\psi_{\mu-2\delta,\nu-2\delta}-
\psi_{\mu+2\delta,\nu-2\delta}\right)\nonumber\\
&=& 0\, .
\end{eqnarray}
For $\mu=0$, we have
\begin{equation} \label{qee3mu0}
\sqrt{|\nu+2\delta|} \psi_{2\delta,\nu+2\delta}
+(\sqrt{|\nu+\delta|}-\sqrt{|\nu-\delta|})
 \psi_{4\delta,\nu}
-\sqrt{|\nu-2\delta|}\psi_{2\delta,\nu-2\delta}= 0\, ;
\end{equation}
for $\mu=\delta$:
\begin{eqnarray} \label{qee3mu1}
&&2(\sqrt{|\nu+2\delta|}+\sqrt{|\nu|}) \psi_{3\delta,\nu+2\delta}+
2(\sqrt{|\nu+2\delta|}-\sqrt{|\nu|}) \psi_{\delta,\nu+2\delta}\nonumber\\
&& +(\sqrt{|\nu+\delta|}-\sqrt{|\nu-\delta|})
\left(3\psi_{5\delta,\nu}+2\psi_{3\delta,\nu}- (1+2\gamma^2\delta^2)\psi_{\delta,\nu}
 \right)\nonumber\\
&&-2(\sqrt{|\nu-2\delta|}+\sqrt{|\nu|})\psi_{3\delta,\nu-2\delta}-
2(\sqrt{|\nu-2\delta|}-\sqrt{|\nu|})\psi_{\delta,\nu-2\delta}
= 0\,,
\end{eqnarray}
for $\mu=2\delta$:
\begin{eqnarray} \label{qee3mu2}
&&(\sqrt{|\nu+2\delta|}+\sqrt{|\nu|}) \psi_{4\delta,\nu+2\delta}-
\sqrt{|\nu|}\psi_{0,\nu+2\delta}\nonumber\\
&& +(\sqrt{|\nu+\delta|}-\sqrt{|\nu-\delta|})
\left(2\psi_{6\delta,\nu}+\psi_{2\delta,\nu}- (1+2\gamma^2\delta^2)\psi_{2\delta,\nu}
 \right)\nonumber\\
&&-(\sqrt{|\nu-2\delta|}+\sqrt{|\nu|})\psi_{4\delta,\nu-2\delta}
+\sqrt{|\nu|}\psi_{0,\nu-2\delta}=0\,,
\end{eqnarray}
and for $\mu=3\delta$:
\begin{eqnarray} \label{qee3mu3}
&&2(\sqrt{|\nu+2\delta|}+\sqrt{|\nu|}) (\psi_{5\delta,\nu+2\delta}-
\psi_{\delta,\nu+2\delta})\nonumber\\
&& +(\sqrt{|\nu+\delta|}-\sqrt{|\nu-\delta|})
\left(5\psi_{7\delta,\nu}+2\psi_{\delta,\nu}- 3(1+2\gamma^2\delta^2)\psi_{3\delta,\nu}
 \right)\nonumber\\
&&-2(\sqrt{|\nu-2\delta|}+\sqrt{|\nu|})(\psi_{5\delta,\nu-2\delta}-
\psi_{\delta,\nu-2\delta})=0\,.
\end{eqnarray}
For values of $\mu$ not an integer multiple of $\delta$ one can derive
similar expressions, but they will not be used in what follows. As
remarked above, strictly speaking, the role of (\ref{qee2}) is
only to select physical wave functions. However, intuitively it
can also be thought of as the quantum evolution equation.  Recall
that $p_c$ can be regarded as an internal time in the classical
theory of the model. Therefore, in the quantum theory, one may
regard $\nu$ as internal time. Then, (\ref{qee2}) can be
interpreted as the quantum Einstein's equation, which evolves the
wave functions in discrete steps of magnitude $2\delta$ along $\nu$.

So far, the parameter $\delta$ ---and hence the size of the
`time-step' --- is arbitrary. In the classical theory, we recover
the Hamiltonian constraint only in the limit $\delta$ goes to
zero. This is because, while the expression (\ref{cotriad}) is
exact for all loops, curvature ${}^oF$  can be expressed in terms
of holonomies only in the limit in which the areas of loops shrink
to zero (see (\ref{F})). In quantum theory, however, the
straightforward limit $\lim_{\delta\to 0} \hat{C}^{(\delta)}$
diverges (for the same reasons as in the isotropic case
\cite{abl}). The viewpoint is that this is occurs because the
limit ignores the quantum nature of geometry, i.e., the fact that
in full quantum geometry, the area operator has a minimum non-zero
eigenvalue. The `correct' quantization of the Hamiltonian
constraint $\ham$ has to take in to account the quantum nature of
geometry.

To do so, note first that the holonomies used in the expression of
$\F$ define quantum states with $\mu=\delta$ and $\nu=\delta$. Using these
states, we can calculate the \emph{quantum} area of each of the
three faces of the elementary cell $\S^2\times {\cal I}$, enclosed
by the three curves ---one along $x$, one along a longitude and
one along the equator of $\S^2$. The area operators defined by
faces $S_{x,\vt}, S_{x,\vp}$ and $S_{\vt,\vp}$ are, respectively,
$2\pi |\hat{p}_b|,\, 2\pi |\hat{p}_b|$ and $\pi |\hat{p}_c|$ (see
(\ref{area})). The quantum geometry states defined by the three
holonomies are eigenstates of these area operators. Furthermore,
they have the same eigenvalue: $\pi\gamma\delta \lp^2$. Now, in the
full theory, we know that the area operator has a minimum non-zero
eigenvalue, $a_o = 2\sqrt{3}\pi\gamma \lp^2$. The viewpoint, as in
\cite{abl}, is that in the calculation of the field strength $\F$,
it is \emph{physically} inappropriate to try to use surfaces of
arbitrarily small areas. The best one can do is to shrink the area
of the loop till it attains this quantum minimum $a_o$. This
implies that we should set $\delta = 2\sqrt{3}$. To summarize, by
using input from the full theory, we conclude that the quantum
Hamiltonian constraint is given by $\hat{C}^{(\delta_o)}$ with $\delta_o =
2\sqrt{3}$.

\emph{Remarks}:

i) Note that the character of the difference equation changes
depending on whether $\mu\ge 4\delta$ or $\mu < 4\delta$. In particular,
for $\mu \ge 4\delta$, the knowledge of the wave function at `times'
$\nu + 2\delta$ and $\nu$ determines only the combination
$\psi_{\mu-2\delta, \nu-2\delta} - \psi_{\mu +2\delta,\nu-2\delta}$ at `time'
$\nu-2\delta$ via (\ref{qee2}). On the other hand, for $\mu =0$ the
wave function at $\nu + 2\delta$ and $\nu$ determines $\psi_{2\delta,
\tau+2\delta}$ completely via (\ref{qee3mu0}). We will return to these
differences in section \ref{s4.3}.

ii) As discussed in \cite{abl,mb8,bd} in detail, one can recover
the Wheeler-DeWitt equation with a specific factor ordering by
taking a systematic limit of quantum Einstein's equation. In the
present model, the difference equation reduces to the following
differential equation for the wave function
$\Psi(p_b,p_c)=\psi_{2p_b/\gamma\lp^2, \, p_c/\gamma\lp^2}$ on the
classical minisuperspace $\M$:
\be\label{wdw} {16\lp^4}\,\left( \sqrt{p_c}\,\f{\partial^2
\Psi}{\partial p_b \partial p_c} + \f{p_b}{4\sqrt{p_c}}\,
\f{\partial^2 \Psi}{\partial p_b^2}+ \frac{1}{2\sqrt{p_c}}
\frac{\partial\Psi}{\partial p_b} \right) - 4 \f{p_b}{\sqrt{p_c}}
\Psi =0 \ee
For $\mu \gg \delta, \nu\gg \delta$, solutions of the discrete equation
(\ref{qee2}) can be approximated by those of this Wheeler DeWitt
equation in a precise sense. In essence, this Wheeler-DeWitt
equation is obtained from (\ref{qee2}) by ignoring quantum
geometry effects, i.e., by an appropriate $\delta \to 0$ limit.  For
the approximation by a differential equation, higher derivatives
of the wave function $\psi$ must be sufficiently small. Thus, the
limit $\delta\to0$ for the equation (\ref{qee2}) exists only under
additional assumptions on the wave function. It does not exist for
the \emph{operator} $\hat{C}_{\rm grav}^{(\delta)}$ by itself.

\subsection{Absence of Singularity}
\label{s4.3}

Recall that the classical singularity occurs at $p_c =0$. The
Wheeler DeWitt equation (\ref{wdw}) is manifestly singular there.
One can multiply it by $\sqrt{p_c}$ and define a new `internal
time, $\bar{p_c} = \ln p_c$ and obtain a regular equation in
variables $p_b,\bar{p}_c$. However, since $\bar{p}_c \rightarrow
-\infty$ at the singularity, this regular equation does not let us
evolve across the singularity. Thus, the overall situation is the
same as in the isotropic model.

To obtain the Wheeler-DeWitt equation from the `fundamental'
difference equation (\ref{qee2}), we had to ignore quantum
geometry effects. Thus, the key question is whether this failure
of the Wheeler-DeWitt equation is an artifact of the approximation
used. Can quantum geometry effects make a qualitative difference
as they do in the isotropic model? We will now argue that the
answer is in the affirmative.

Let us analyze the `evolution' given by the difference equation
(\ref{qee2}). Does this evolution stop at $\nu =0$? As in the
isotropic case \cite{mb3,mb7}, we will use the quantum constraint
(\ref{qee2}) as a recurrence relation starting at positive $\nu$
and evolve toward smaller values. However, now a new twist arises
because for generic values $\mu \ge 4\delta$, (\ref{qee2}) determines
only the difference $\psi_{\mu+2\delta,\nu-2\delta}
-\psi_{\mu-2\delta,\nu-2\delta}$ as a function of the initial values of
the wave function. Therefore, quantum Einstein's equation has to
be supplemented with an appropriate boundary condition. From
(\ref{qee2}) which holds for $\mu \ge 4\delta$, one might conclude
that, even if we restrict ourselves to the `lattice' $\mu = n\delta$,
one would have to specify the wave function at $\mu=0, \delta, 2\delta,
3\delta$ at each `time step'. However, as remarked at the end of
section \ref{s4.2}, the form of the difference equations for $\mu
< 4\delta$ is different. An examination of (\ref{qee3mu0}) and
(\ref{qee3mu1}) reveals that it is in fact sufficient to specify
the wave function $\psi_{\mu,\nu}$ just at $\mu=0$ and $\mu=\delta$ at
each `time-step'. This is a mathematically viable choice of the
boundary condition and heuristically it corresponds to providing
data the `horizon'. It will turn out that the issue of the
resolution of singularity is insensitive to the precise choice of
the boundary condition. We will therefore postpone the discussion
of physically appropriate choices until after the main discussion.

Let us then start at some finite positive $\nu$ and carry out a
backward evolution using (\ref{qee2}). In contrast to the
Wheeler-DeWitt equation (\ref{wdw}), the coefficients in the
difference equation (\ref{qee2}) are always regular. At first one
might think that this regularity by itself would be sufficient to
ensure that the evolution would be well-defined across the
singularity. Note however that this need not be the case. For, the
backward evolution continues only as long as the coefficient of
the wave function at $\nu-2\delta$ is non-zero. The key question
therefore is whether this coefficient vanishes. The issue is
subtle. For instance,  the coefficients in the non-self-adjoint
operator $(\hat{C}^{\delta})^\dag$ are also non-singular. However, if
one uses it in place of $\hat{C}^{\delta}_{\rm Ham}$, one finds that
the coefficient vanishes and one can not evolve across the
singularity.

For our quantum Einstein's equation (\ref{qee2}), this coefficient
is given by $\sqrt{|\nu-2\delta|}+\sqrt{|\nu|}$. By inspection,
\emph{it never vanishes}. Therefore, the backward quantum
evolution remains well-defined and determines the wave function
not only for $\nu >0$  but also in the new region with $\nu\le 0$.
In this precise sense, the classical black hole singularity can be
traversed using quantum evolution and thus ceases to be a boundary
of space-time.

Next, recall from section \ref{s4.1} that we could have begun with
the expression (\ref{hami}) of the Hamiltonian constraint in the
classical theory and then proceeded with quantization. What would
be the status of the singularity resolution with this choice? The
procedure to construct the quantum Hamiltonian constraint would be
identical. However, the final result would be slightly different:
$\gamma^2$ in (\ref{qee2}) would be replaced by $1+\gamma^2$. That
is, in the new quantum Einstein equation, the coefficient of
$\psi_\mu,\nu$ in (\ref{qee2}) would be modified and the equation
would become:
\begin{eqnarray} \label{qee4}
&&2\delta(\sqrt{|\nu+2\delta|}+\sqrt{|\nu|}) (\psi_{\mu+2\delta,\nu+2\delta}-
\psi_{\mu-2\delta,\nu+2\delta})\nonumber\\
&& +(\sqrt{|\nu+\delta|}-\sqrt{|\nu-\delta|})
\left((\mu+2\delta)\psi_{\mu+4\delta,\nu}- (1+2(\gamma^2+1)
\delta^2)\mu\psi_{\mu,\nu}+
(\mu-2\delta)\psi_{\mu-4\delta,\nu}\right)\nonumber\\
&&+2\delta(\sqrt{|\nu-2\delta|}+\sqrt{|\nu|})(\psi_{\mu-2\delta,\nu-2\delta}-
\psi_{\mu+2\delta,\nu-2\delta})\nonumber\\
&=& 0
\end{eqnarray}
(The same replacement holds for $\mu < 4\delta$.) Since the
coefficient of $\psi_{\mu-2\delta,\nu-2\delta}- \psi_{\mu+2\delta,\nu-2\delta}$ is
unaltered, one would again be able to `evolve' across the
singularity. Thus, the conclusion is robust within quantum
geometry.%
\footnote{However, the relation between the wave function and the
Wheeler--DeWitt approximation is less direct with this form of the
constraint.}

To conclude, we will briefly return to the question of the
boundary condition that are needed to make the evolution
well-defined even away from the singularity. As noted earlier, in
the backward evolution given by (\ref{qee3mu0}),\,
$\psi_{2\delta,\nu_o}$ is determined by values of the wave function at
$\nu>\nu_o$, so long as for $\nu_o\not=0$. Furthermore,
(\ref{qee2}) then determines the wave function at $\mu=(2+4n)\delta$
for all $n\in{\mathbb N}$ and at that $\nu_o$. One could now use
pre-classicality arguments \cite{mb4} (i.e.\ require that the wave
function not oscillate on small scales) at large $\mu$ to choose
the boundary values, eliminating entirely the need for specifying
boundary conditions. While this could be reasonable in
semiclassical regimes, for small $|\nu|$ it would be questionable
to use pre-classicality in the $\mu$-direction even if a
pre-classical solution exist (which is not guaranteed, see e.g.\
the analysis in \cite{ck,gd}). More importantly, the slice $\nu=0$
is special because $\psi_{2\delta,0}$ is not determined through
(\ref{qee3mu0}) or otherwise. At this point, the condition
(\ref{qee3mu0}) evaluated at $\nu=-2\delta$ gives instead a condition
for previous values of the wave function, translating into one
condition for the initial values. At $\nu=0$, however,
$\psi_{2\delta,0}$ has to be specified as boundary value in addition
to $\psi_{0,0}$ because, unlike similar situations in isotropic
models, it does not drop out of the evolution: it is needed in the
recurrence (\ref{qee2}) for $\mu=4\delta$, $\nu=2\delta$. There is thus no
reduction in conditions on the wave function from the fact that
$\psi_{2\delta,0}$ drops out of (\ref{qee3mu0}); a condition is simply
transferred from boundary values to initial values. Note however
that, even though the line $\nu=0$ (corresponding to the classical
singularity) does show special behavior, evolution does not break
down there; it is only the boundary value problem which changes
character.

While such a boundary value problem is mathematically
well-defined, the resulting theory is not necessarily physically
correct. For, in a physically interesting theory one would expect
that, well away from the singularity, there should exist
semi-classical solutions which are peaked on the classical
trajectories. From Fig.~1 it is clear that the semi-classical
solution peaked on the trajectory labelled by mass $m$ would be
sharply peaked at $\nu= \pm 4m^2$ on the line $\mu=0$, whence
semi-classical states peaked at different classical solutions will
have quite different forms near $\mu=0$. Therefore, the theory
obtained by simply fixing the wave function on (or near) $\mu=0$
is not likely to admit a sufficiently rich semi-classical sector.
Indeed, it is not clear that any boundary condition at (or near)
$\mu=0$ will be physically viable. Rather, one may have to impose
the boundary condition at $\mu=\infty$, e.g., by requiring that
the wave function should vanish there.  Indeed, the form of
dynamical trajectories of Fig.~1 implies that \emph{every}
semi-classical state would share this property. Moreover, the
corresponding condition does hold in the closed isotropic model
with a massless scalar field as source \cite{aps2}. As in that
case, one would expect that these semi-classical states would span
a dense subspace in the physical Hilbert space. If so, requiring
that the wave functions vanish at $\mu=\infty$ would be physically
justified. Heuristics suggest that this strategy is viable for the
Wheeler-DeWitt equation (\ref{wdw}). However, detailed numerical
analysis is necessary to establish that the difference equations
(\ref{qee2}) or (\ref{qee4}) admits solutions satisfying this
condition (or a suitable modification thereof) for all $\nu$ and
that its imposition makes the evolution unambiguous.

\section{Discussion}
\label{s5}

Results of the last two sections support a general scenario that
has emerged from the analysis of singularities in quantum
cosmology. It suggests that the classical singularity does not
represent a final frontier; the \emph{physical} space-time does
not end there. In the Planck regime, quantum fluctuations do
indeed become so strong that the classical description breaks
down. The space-time continuum of classical general relativity is
replaced by discrete quantum geometry which remains regular during
the transition through what was a classical singularity. Certain
similarities between the Kantowski-Sachs model analyzed here and a
cosmological model which has been studied in detail \cite{aps}
suggest that there would be a quantum bounce to another large
classical region. If this is borne out by detailed numerical
calculations, one would conclude that quantum geometry in the
Planck regime serves as a bridge between two large classical
regions. Space-time may be much larger than general relativity has
had us believe.

However, as indicated at the end of section \ref{s4.2} significant
numerical work is still needed before one can be certain that this
scenario is really borne out in the model. Moreover, this is a
highly simplified model. It is important to check if the
qualitative conclusions remain robust as more and more realistic
features are introduced. First, one should extend the analysis so
that the space-time region outside the horizon is also covered. A
second and much more important challenge is incorporation of an
infinite number of degrees of freedom by coupling the model, e.g.,
to a spherically symmetric scalar field. First steps along these
lines have been taken \cite{mbss,hw} and one can see that the
evolution still extends beyond the classical singularity
\cite{mb9}. But a comprehensive treatment still remains a distant
goal.

\section*{Acknowledgements:}

This work was supported in part by NSF grants PHY-0090091, and
PHY-0354932, the Alexander von Humboldt Foundation, the C.V. Raman
Chair of the Indian Academy of Sciences and the Eberly research
funds of Penn State.

\end{document}